\documentclass[12pt]{article}
\usepackage{amsmath}
\date{}
\begin{document}

\title{Separation of Coupled Systems of Schrodinger Equations  
by Darboux transformations}

\author{Mayer Humi\\
Department of Mathematical Sciences\\
Worcester Polytechnic Institute\\
100 Institute Road\\
Worcester, MA  0l609}

\maketitle
\thispagestyle{empty}

\begin{abstract}
Darboux transformations in one independent variable have found 
numerous applications in various field of mathematics and physics.
In this paper we show that the extension of these transformations to
two dimensions can be used to decouple systems of Schrodinger 
equations and provide explicit representation for three classes of such 
systems.  We show also that there is an elegant relationship between 
these transformations and analytic complex matrix functions.
\end{abstract}
\vspace{1.in}

\noindent MSC(2010):\,\, 35J10, 35J47, 81Q80

\newpage

\section{Introduction}

For many years Darboux transformations in one independent variable
have found numerous applications in various field of mathematics and physics
[1-17]. (Refs.  [1,2] contain an extensive list of references). In 
particular the factorization method and it generalizations [13-14] 
which have been instrumental in many physical applications (including 
SUSY QM [15] ) has its roots based on these transformations. Recently 
however these transformations were generalized and applied to systems 
of nonlinear equations such as the KdV hierarchy and soliton solutions 
[2,5,7]. In addition various applications of this method in geometry were 
worked out and form an important ongoing research area [2]. Extensions of 
the method to multidimensional oriented Riemann manifolds [8], time dependent 
potentials [9,16] and shape invariant potentials [17] have appeared in 
the literature.

It is surprising that in spite of of this extensive research effort the 
theory and applications of these transformations in two variables and its 
elegant relationship to complex analytic function theory has not been 
recognized widely in the literature.[4] 

Here is a short overview of Darboux transformations for Schrodinger 
equation in one variable. (In the following we assume that all functions
under consideration are smooth).

We say that the solutions of two Schrodinger equations with different
potentials $u(x), v(x)$ i.e.
\begin{equation}
\label{1.1}
\phi^{\prime\prime} = ( u(x)+ \lambda) \phi
\end{equation}
\begin{equation}
\label{1.2}
\psi^{\prime\prime} = ( v(x) + \lambda) \psi.
\end{equation}
are related by a Darboux transformation if there exist $A(x),B(x)$
so that 
\begin{equation}
\label{1.3}
\psi = \left[A(x) + B(x) \frac{\partial}{\partial x}\right]\phi(x).
\end{equation}
Letting $B(x) = 1$  one can easily show that in order for eqs
(\ref{1.1}), (\ref{1.2}) to be related by the transformation
(\ref{1.3}) $A(x), u(x), v(x)$ must satisfy;
\begin{equation}
\label{1.4}
A^{\prime\prime} + u^\prime + A(u-v) = 0
\end{equation}
\begin{equation}
\label{1.5}
2A^\prime + u - v = 0
\end{equation}
Eliminating  $(u-v)$ between these equations and integration yields
\begin{equation}
\label{1.6}
A^\prime - A^2 + u = -\nu
\end{equation}
where  $\nu$ is an integration constant.  Eq. (\ref{1.6}) is a
Riccati equation which can be linearized by the transformation  $A =
-\zeta^\prime/\zeta$  which leads to
\begin{equation}
\label{1.7}
\zeta^{\prime\prime} = (u(x) + \nu)\zeta.
\end{equation}
Thus $\zeta$ is an eigenfunction of the original eq. (\ref{1.2}) with
$\lambda = \nu$.  From (\ref{1.5}) we now infer that
\begin{equation}
\label{1.8}
v = u - 2(ln\zeta)^{\prime\prime}
\end{equation}
i.e. a Darboux transformation changes the potential function  $u(x)$
by  $\Delta u = -2(ln\zeta)^{\prime\prime}$  where  $\zeta$ is an
arbitrary eigenfunction of (\ref{1.1}).

From a more general point of view we note that Darboux transformation for 
equations in real variables form a special case of the Laplace-Darboux 
transformations for the complex variable linear equation [2]
\begin{equation}
\label{1.11}
{\bf\Psi}_{z{\bar z}} +a{\bf\Psi}_{z}+b{\bf\Psi}_{\bar z}+{\bf V}\bf{\Psi} = \bf{0}
\end{equation}
(where $z=x+iy$).

Our objective in this paper is to introduce a novel application of these 
transformations to decouple systems of partial differential equations in 
two dimensions of the form
\begin{equation}
\label{1.9}
\nabla^2{\bf \Psi} +{\bf V}(x,y)\bf{\Psi} = \bf{0}
\end{equation}
where ${\bf\Psi}^{T}= (\psi_1(x,y),\psi_2(x,y))$ and ${\bf V}(x,y)$ is a 
$2\times 2$ matrix. Such systems appear frequently in many applications
of classical and modern physics [4,6]. 

We note that the more general form of this equation
\begin{equation}
\label{1.10}
{\bf A}(x,y)\nabla^2{\bf\Psi} +{\bf V}(x,y)\bf{\Psi} = \bf{0}
\end{equation}
can be transformed into (\ref{1.9}) if ${\bf A}(x,y)$ is (locally) invertible.

The plan of the paper is as follows: In Sec 2 we derive the basic equations
that constrain Darboux transformations in two dimensions.
In Secs. $3,4,5$ we solve these equations explicitly for three 
different classes of the potential matrix ${\bf V}$ and
end up in Sec. $6$ with summary and conclusions. 

\setcounter{equation}{0}
\section{Darboux transformations in two Dimensions.}

We shall say that two systems of partial differential equations (PDEs)
in two independent variables
\begin{equation}
\label{2.1}
\nabla^2{\bf\Phi} +  {\bf U}(x,y){\bf\Phi} = \bf{0}
\end{equation}
\begin{equation}
\label{2.2}
\nabla^2{\bf\Psi} +  {\bf V}(x,y){\bf\Psi} = \bf{0}
\end{equation}
are related by a Darboux transformation if there exist $2\times 2$
nonsingular matrices with smooth function entries ${\bf C}_1(x,y),
{\bf C}_2(x,y), {\bf C}_3(x,y)$ so that their solutions satisfy 
\begin{equation}
\label{2.3}
{\bf\Psi} = \left[{\bf C}_1(x,y) + {\bf C}_2(x,y) \frac{\partial}{\partial x}+ {\bf C}_3(x,y)\frac{\partial}{\partial y}\right]{\bf\Phi}.
\end{equation}

For brevity we drop in the following the dependence of the various
functions on the independent variables.

Using eq. (\ref{2.3}) to substitute for ${\bf\Psi}$ in eq. (\ref{2.2}) and
eliminating the higher order derivatives of ${\bf\Phi}$,
$\nabla^2{\bf\Phi}$ and $\frac{\partial^{2}{\bf\Phi}}{\partial {y}^{2}}$ using eq. (\ref{2.1}) we obtain
\begin{eqnarray}
\label{2.4}
2\left[ {\frac {\partial {\bf C}_2 }{\partial y}} -{
\frac {\partial {\bf C}_3 }{\partial x}}  \right] {\frac {
\partial ^{2}{\bf\Phi}}{\partial {x}^{2}}} +  
2\left[{ \frac {\partial {\bf C}_2 }{\partial y}} +{\frac { \partial {\bf C}_3  }{\partial x}} \right] {\frac {\partial 
^{2}{\bf\Phi}}{\partial x\partial y}} +   \nonumber
\end{eqnarray}
\begin{eqnarray}
\left\{ \nabla^2 {\bf C}_3   + 2\,{\frac {\partial {\bf C}_1  }{
\partial y}} - {\bf C}_3\bf{U} +\bf{V}{\bf C}_3 
\right\} {\frac {\partial {\bf\Phi} }{\partial y}} + 
\left\{ \nabla^2{\bf C}_2+2\,{\frac {\partial {\bf C}_1 }{\partial x}} 
-{\bf C}_2\bf{U}+\bf{V}{\bf C}_2 \right\} {\frac {\partial {\bf \Phi} }{\partial x}} +  \nonumber
\end{eqnarray}
\begin{eqnarray}
\left\{ \nabla^2{\bf C}_1- 2\, { \frac {\partial {\bf C}_3}{\partial y}}
{\bf U} - {\bf C}_2 {\frac {\partial {\bf U} }
{\partial x}} - {\bf C}_3  {\frac {\partial {\bf U} }{\partial y}}
-{\bf C}_1\bf{U} +\bf{V}{\bf C}_1\right\} {\bf\Phi}  = {\bf 0}
\end{eqnarray}

To satisfy this equation we treat ${\bf\Phi}$ and its derivatives as 
independent variables and let their coefficients be zero. This leads then 
to the following system of equations.

\begin{equation}
\label{2.5}
{\frac {\partial {\bf C}_2 }{\partial x}}-{\frac {\partial {\bf C}_3 }
{\partial y}}= {\bf 0}
\end{equation}
\begin{equation}
\label{2.6}
{\frac {\partial {\bf C}_2 }{\partial y}}+{\frac {\partial {\bf C}_3 }
{\partial x}}= {\bf 0}
\end{equation}
\begin{equation}
\label{2.7}
\nabla^2 {\bf C}_2+2{\frac {\partial {\bf C}_1 }{\partial x}} 
-{\bf C}_2\bf{U}+\bf{V}{\bf C}_2 = {\bf 0}
\end{equation}
\begin{equation}
\label{2.8}
\nabla^2{\bf C}_3 + 2\,{\frac {\partial {\bf C}_1 }{ \partial y}} 
-{\bf C}_3\bf{U}+\bf{V}{\bf C}_3= {\bf 0}
\end{equation}
\begin{equation}
\label{2.9}
\nabla^2{\bf C}_1 -2\,{ \frac {\partial {\bf C}_3}{\partial y}}\bf{U} - 
{\bf C}_2{\frac {\partial \bf{U} }{\partial x}} - 
{\bf C}_3{\frac {\partial \bf{U} }{\partial y}}-{\bf C}_1\bf{U}+
\bf{V}{\bf C}_1 = {\bf 0}
\end{equation}
We observe that eq. (\ref{2.9}) can be rewritten in a symmetric form
in ${\bf C}_2,{\bf C}_3$ in view of eq.(\ref{2.5}).

Eqs.(\ref{2.5}),(\ref{2.6}) are Cauchy-Riemann equations for 
${\bf C}_2,{\bf C}_3$. Hence the corresponding entries in this matrices
must be harmonic conjugates and the entries of 
\begin{equation}
\label{2.10}
{\bf D}={\bf C}_2+i{\bf C}_3 
\end{equation}
are analytic. In view of this fact $\nabla^2{\bf C}_2=\nabla^2{\bf C}_3=
{\bf 0}$ and eqs.  (\ref{2.7}),(\ref{2.8}) simplify to
\begin{equation}
\label{2.11}
2\,{\frac {\partial {\bf C}_1 }{\partial x}} -{\bf C}_2\bf{U}+
\bf{V}{\bf C}_2 = {\bf 0}, \;\;
2\,{\frac {\partial {\bf C}_1 }{\partial y}} -{\bf C}_3\bf{U}+
\bf{V}{\bf C}_3= {\bf 0}.
\end{equation}
From (\ref{2.11}) we have
\begin{equation}
\label{2.12}
\bf{V}= [{\bf C}_2\bf{U}-2\,{\frac {\partial {\bf C}_1 }{\partial x}}]
{\bf C}_2^{-1}, \;\;\;
\bf{V}= [{\bf C}_3\bf{U}-2\,{\frac {\partial {\bf C}_1 }{\partial y}}]
{\bf C}_3^{-1}.
\end{equation}
Hence
\begin{equation}
\label{2.13}
2\left[{\frac {\partial {\bf C}_1 }{\partial x}}{\bf C}_2^{-1}- {\frac {\partial {\bf C}_1 }{\partial y}}{\bf C}_3^{-1}\right] =
{\bf C}_2{\bf U}{\bf C}_2^{-1}-{\bf C}_3{\bf U}{\bf C}_3^{-1} \;\;
\end{equation}

Using (\ref{2.12}) to eliminate $\bf{V}$ from (\ref{2.9}) leads to
\begin{equation}
\label{2.14}
\nabla^2{\bf C}_1 -2\,{ \frac {\partial {\bf C}_3}{\partial y}}\bf{U} - 
{\bf C}_2{\frac {\partial \bf{U} }{\partial x}} - 
{\bf C}_3{\frac {\partial \bf{U} }{\partial y}}-{\bf C}_1\bf{U}+
[{\bf C}_3\bf{U}-2\,{\frac {\partial {\bf C}_1 }{\partial y}}]
{\bf C}_3^{-1}{\bf C}_1 = {\bf 0}
\end{equation}

Assuming that ${\bf C}_2, {\bf C}_3$ were chosen already to satisfy 
(\ref{2.5})- (\ref{2.6}) the system (\ref{2.13})-(\ref{2.14}) consists 
of eight coupled nonlinear equations for the entries of ${\bf C}_1$ 
and ${\bf U}$. However in order to decouple the system (\ref{2.2}) by the 
transformation (\ref{2.3}) the resulting ${\bf U}$ must be diagonal or upper 
(lower) triangular.  In the following we assume that the desired form of the 
matrix ${\bf U}$ is diagonal viz.
\begin{eqnarray}
\label{3.2}
{\bf U} = \left(\begin{array}{cc}
u_{11}(x,y) &0  \\
0 &u_{22}(x,y) \\
\end{array}
\right).
\end{eqnarray}
Under this constraint the system (\ref{2.13})-(\ref{2.14})
is an over determined system of eight coupled nonlinear equations in 
six (unknown) functions. 

To derive solutions for these equations our strategy will be as follows: 
First we choose 
a proper form for the entries of the matrices ${\bf C}_2$ and ${\bf C}_3$. 
Then we solve (\ref{2.13})-(\ref{2.14}) (under some restrictions) for 
${\bf C}_1$ and ${\bf U}$. Finally we obtain ${\bf V}$  from (\ref{2.12}).  

In the next three sections we provide solutions and explicit 
examples for this decoupling procedure for
different choices of the matrices ${\bf C}_2$ and ${\bf C}_3$ 

\setcounter{equation}{0}
\section{${\bf C}_2$ and ${\bf C}_3$ are Matrices with Constant Entries}

When ${\bf C}_2$ and ${\bf C}_3$ are matrices with real constants entries viz.
\begin{eqnarray}
\label{3.15}
{\bf C}_2 = \left(\begin{array}{cc}
d_{11} &d_{12}  \\
d_{21} &d_{22} \\
\end{array}
\right),\,\,\,
{\bf C}_3 = \left(\begin{array}{cc}
e_{11} &e_{12}  \\
e_{21} &e_{22} \\
\end{array}
\right).
\end{eqnarray}
equations (\ref{2.5}), (\ref{2.6}) are satisfied by default.
We also assume that ${\bf C}_1$ is of the form
\begin{eqnarray}
\label{3.3}
{\bf C}_1 = \left(\begin{array}{cc}
c_{11}(x,y) &c_{12}(x,y)  \\
c_{21}(x,y) &c_{22}(x,y) \\
\end{array}
\right).
\end{eqnarray}
 
In this setting eq. (\ref{2.13}) leads to a coupled system of equations 
which can be simplified if we assume that
\begin{equation}
\label{3.16}
e_{21}=\frac{d_{21}e_{11}}{d_{11}},\,\,\, 
e_{12}=\frac{d_{12}e_{22}}{d_{22}}
\end{equation}
The resulting equations for the entries of the matrix ${\bf C}_1$ can be 
solved (after some algebra) and we find that
the general solution for these entries is
\begin{equation}
\label{3.17}
c_{11}=F_{11}(w_1),\,\,\, c_{12}=F_{12}(w_2),\,\,\,
c_{21}=F_{21}(w_1),\,\,\, c_{22}=F_{22}(w_2),
\end{equation}
where
$$
w_1=d_{11}x+e_{11}y, \,\,\, w_2=e_{22}y+d_{22}x,
$$
and $F_{ij}$ are smooth functions of the indicated variable. To proceed 
we assume a special form of the functions $F_{ij}$,\, 
${i,j=1,2}$
\begin{eqnarray}
\label{3.18}
&&F_{11}=h_{11}(d_{11}x+ e_{11}y),\,\,\, 
F_{12}=\frac{h_{22}d_{12}(d_{22}x+ e_{22}y)}{d_{22}},  \\ \notag
&&F_{21}=\frac{h_{11}d_{21}(d_{11}x+ e_{11}y)}{d_{11}},\,\,\, 
F_{22}=h_{22}(d_{22}x+ e_{22}y))
\end{eqnarray}
where $h_{ij}$ are constants. Using these expressions to evaluate the 
right hand side of 
\begin{equation}
\label{3.9}
{\bf E}=\nabla^2{\bf C}_1 -2\,{\frac {\partial {\bf C}_1 }{\partial y}}
{\bf C}_3^{-1}{\bf C}_1. 
\end{equation}
We find:
\begin{eqnarray}
\label{3.19}
{\bf E} = \left(\begin{array}{cc}
-2h_{11}^2(d_{11}x+ e_{11}y) &\frac{-2h_{22}^2d_{12}(d_{22}x+ e_{22}y)}{d_{22}}  \\ \\
\frac{-2h_{11}^2d_{21}(d_{11}x+ e_{11}y)}{d_{11}} & -2h_{22}^2(d_{22}x+ e_{22}y))\\ 
\end{array}
\right).
\end{eqnarray}
Substituting this result in (\ref{2.14}) and using (\ref{3.16}) we obtain only 
two independent equations for $u_{11}$ and $u_{22}$
\begin{equation}
\label{3.20}
d_{11}\frac{\partial u_{11}(x, y)}{\partial x} +
e_{11}\frac{\partial u_{11}(x, y)}{\partial y}+2h_{11}^2(d_{11}x+ e_{11}y) =0,
\end{equation}
\begin{equation}
\label{3.21}
d_{22}\frac{\partial u_{22}(x, y)}{\partial x} +
e_{22}\frac{\partial u_{22}(x, y)}{\partial y}+2h_{22}^2(d_{22}x+ e_{22}y) =0.
\end{equation}
The general solution to these equations is
\begin{equation}
\label{3.22}
u_{11}=\frac{h_{11}^2}{2}\frac{e_{11}^2(e_{11}^2-d_{11}^2)x^2+
d_{11}^2(d_{11}^2-e_{11}^2)y^2-
2d_{11}e_{11}(d_{11}^2+e_{11}^2)xy}{d_{11}^2e_{11}^2}+H_1(\frac{d_{11}y-e_{11}x}{d_{11}})
\end{equation}
\begin{equation}
\label{3.23}
u_{22}=\frac{h_{22}^2}{2}\frac{e_{22}^2(e_{22}^2-d_{22}^2)x^2+
d_{22}^2(d_{22}^2-e_{22}^2)y^2
-2d_{22}e_{22}(d_{22}^2+e{22}^2)xy}{d_{22}^2e_{22}^2}+
H_2(\frac{d_{22}y-e_{22}x)}{d_{22}})
\end{equation}

where $H_1$ and $H_2$ are arbitrary (smooth) functions of the indicated 
variables.

The general form of the matrix ${\bf V}$ (from (\ref{2.12})) is
\begin{eqnarray}
\label{3.24}
{\bf V} = \frac{1}{det({\bf C}_2)}\left(\begin{array}{cc}
d_{11}d_{22}(u_{11}-2h_{11})-d_{12}d_{21}(u_{22}-2h_{22})
&d_{11}d_{12}(u_{22}-u_{11}+2(h_{11}-h_{22})) \\ \\
d_{22}d_{21}(u_{11}-u_{22}+2(h_{22}-h_{11})) 
& d_{21}d_{12}(u_{11}-2h_{11})-d_{11}d_{22}(u_{22}-2h_{22})
\end{array}
\right).
\end{eqnarray}
where $det({\bf C}_2)$ is the determinant of ${\bf C}_2$.

{\bf Example}: Let the matrices $C_2$ and $C_3$ be chosen as
\begin{eqnarray}
\label{5.1}
{\bf C}_2 = \left(\begin{array}{cc}
1 &1 \\
1 &-1 \\
\end{array}
\right),\,\,\,
{\bf C}_3 = \left(\begin{array}{cc}
2 &1  \\
2 &-1 \\
\end{array}
\right).
\end{eqnarray}
Furthermore let $h_{11}=h_{22}=1$ and (for simplicity) $H_1=H_2=0$. 
The potential matrix $V$ for the coupled system becomes
\begin{eqnarray}
\label{5.2}
{\bf V} = -\frac{1}{4}\left(\begin{array}{cc}
-3x^2+\frac{3}{4}y^2+9xy+8
& -3x^2+\frac{3}{4}y^2+xy \\ \\
3x^2-\frac{3}{4}y^2-xy
&3x^2-\frac{3}{4}y^2-9xy-8
\end{array}
\right).
\end{eqnarray}
Thus the coupled system represents two coupled two
dimensional oscillators. For the resulting uncoupled system we have
\begin{equation}
\label{5.3}
u_{11}=\frac{1}{2}\left(3x^2-\frac{3}{4}y^2-5xy\right),\,\,\,
u_{22}= -2xy.
\end{equation}

\setcounter{equation}{0}
\section{The Matrix $D=z^nC$, where $C$ has Constant Entries}

In this case we assume that the matrices ${\bf C_2}$ and ${\bf C_3}$ 
are of the form
\begin{eqnarray}
\label{3.25}
{\bf C}_2 = Re(z^n)\left(\begin{array}{cc}
d_{11} &d_{12}  \\
d_{21} &d_{22} \\
\end{array}
\right),\,\,\,
{\bf C}_3 = Im(z^n)\left(\begin{array}{cc}
e_{11} &e_{12}  \\
e_{21} &e_{22} \\
\end{array}
\right).
\end{eqnarray}
where $e_{ij},d_{ij}$ are real and $Re(z^n),Im(z^n)$ are the real and 
imaginary parts of $z^n,\, 0\, <\, n  $.
(Similar treatment can be made if we replace $z^n$ by  $iz^n$). In view of 
(\ref{2.5})-(\ref{2.6}) we must have also $e_{ij}=d_{ij}$.

Since in polar coordinates $Re(z^n)=r^n\cos(n\theta)$ and 
$Im(z^n)=r^n\sin(n\theta)$ it is expedient to work in this coordinate 
system. 

It is easy to see that in this setting the right hand side of (\ref{2.13}) 
is zero and we obtain the following equation for the the elements $c_{ij}$
of ${\bf C}_1$
\begin{equation}
\label{3.26}
\sin((n-1)\theta)\frac{\partial c_{ij}}{\partial r}
-\frac{\cos((n-1)\theta)}{r}\frac{\partial c_{ij}}{\partial \theta} =0,\,\,\,
i,j=1,2.
\end{equation}
The general solution of this equation is $c_{ij}=F_{ij}(w)$ where
\begin{equation}
\label{3.27}
w(r,t)=\frac{cos((n-1)\theta)}{r^{n-1}}
\end{equation}
and $F_{ij}$ are smooth functions of $w$. However in order to obtain
a consistent and decoupled system of equations from (\ref{2.14}) we shall 
assume the following special form of ${\bf C}_1$
\begin{eqnarray}
\label{3.28}
{\bf C}_1 = \left(\begin{array}{cc}
F_{11} &\frac{d_{12}F_{22}}{d_{22}}  \\
\frac{d_{21}F_{11}}{d_{11}} &F_{22} \\
\end{array}
\right).
\end{eqnarray}
With these choices for the elements of ${\bf C}_1$ the matrix ${\bf E}$
has only two independent entries $E_{11}$ and $E_{22}$. (To derive this 
result we used (\ref{3.27})).
\begin{equation}
\label{3.29}
{\bf E}_{ii}=\frac{1}{r^{2n}}[F_{ii}^{\prime}+
\frac{1}{d_{ii}}F_{ii}^2]^{\prime}=\frac{h_{ii}}{r^{2n}}
\end{equation}
where primes denote differentiation with respect to $w$.
The matrix ${\bf E}$ takes the following form:
\begin{eqnarray}
\label{3.30}
{\bf E} =\frac{1}{r^{2n}} \left(\begin{array}{cc}
h_{11} &\frac{d_{12}h_{22}}{d_{22}}  \\
\frac{d_{21}h_{11}}{d_{11}} &h_{22} \\
\end{array}
\right).
\end{eqnarray}
where $h_{ii}=h_{ii}(w)$.
Substituting these results in (\ref{2.14}) and using (\ref{3.27}) we 
find that this system also have only two independent equations
\begin{equation}
\label{3.31}
r^n\cos((n-1)\theta)\frac{\partial u_{ii}}{\partial r}+
r^{n-1}\sin((n-1)\theta)\frac{\partial u_{ii}}{\partial \theta}+
2nr^{n-1}\cos((n-1)\theta)u_{ii}=\frac{h_{ii}}{d_{ii}r^{2n}}. 
\end{equation}
If we choose $F_{ii}$ so that $h_{ii}=0$ then the general solution of
(\ref{3.31}) is
\begin{equation}
\label{3.32}
u_{ii}= \frac{1}{r^{2n}}H_{ii}\left(\frac{\sin((n-1)\theta)}{r^{n-1}}\right)
\end{equation}
where $H_{ii}$ are smooth functions of the indicated variable. 
The general form of the matrix ${\bf V}$ in this case is
\begin{equation}
\label{3.33}
V_{11}= \frac{1}{det(D)}\left[d_{11}d_{22}u_{11}-d_{12}d_{21}u_{22}+ 4r^{-2n}(d_{22}F_{11}^{\prime}-\frac{d_{21}d_{12}}{d_{22}}F_{22}^{\prime})\right]
\end{equation}
\begin{equation}
\label{3.34}
V_{12}= \frac{1}{det(D)}\left[d_{11}d_{12}u_{11}-d_{11}d_{12}u_{22}+ 4r^{-2n}(d_{12}F_{11}^{\prime}-\frac{d_{11}d_{12}F_{22}^{\prime}}{d_{22}})\right]
\end{equation}
\begin{equation}
\label{3.35}
V_{21}= \frac{1}{det(D)}\left[d_{21}d_{22}u_{11}-d_{21}d_{22}u_{22}+ 4r^{-2n}(\frac{d_{21}d_{22}F_{11}^{\prime}}{d_{11}}-d_{21}F_{22}^{\prime})\right]
\end{equation}
\begin{equation}
\label{3.36}
V_{22}= \frac{1}{det(D)}\left[d_{21}d_{12}u_{11}-d_{11}d_{22}u_{22}+ 4r^{-2n}(\frac{d_{21}d_{12}F_{11}^{\prime}}{d_{11}}-d_{11}F_{22}^{\prime})\right]
\end{equation}
where $det(D)=d_{11}d_{22}-d_{12}d_{21}$.

We present two examples.

{\bf Example}: Let $n=1$ and 
\begin{eqnarray}
\label{5.4}
{\bf C} = \left(\begin{array}{cc}
1 & 1 \\
-1 &1 \\
\end{array}
\right).
\end{eqnarray}
Furthermore to satisfy (\ref{3.29}) we let $F_{11}=F_{22} = \frac{1}{r}$. 
For these choices we find that
\begin{equation}
\label{5.5}
u_{11}=\frac{A}{r^2},\,\,\, u_{22}=\frac{B}{r^2}
\end{equation}
where $A,B$ are constants. The corresponding matrix potential for the
coupled system is
\begin{eqnarray}
\label{5.6}
{\bf V} = \left(\begin{array}{cc}
\frac{A+B}{2r^2}-\frac{4}{r^6}
& \frac{A-B}{2r^2} \\ \\
-\frac{A-B}{2r^2}
&-\frac{A+B}{2r^2}+\frac{4}{r^6}
\end{array}
\right).
\end{eqnarray}
We observe that in this case the potentials are radial (no dependence on 
$\theta$).

{\bf Example}: Let $n=2$ and choose $C$ as in the previous example. 
To satisfy (\ref{3.29}) with $h_{ii}=0$ we choose 
$F_{11}=F_{22} = \frac{r}{r+\cos\theta}$ and let 
\begin{equation}
\label{5.7}
H_{11}=\frac{A\sin\theta}{r},\,\,\, H_{22}=\frac{B\sin\theta}{r}.
\end{equation}
With these choices we have
\begin{equation}
\label{5.8}
u_{11}=\frac{A\sin\theta}{r^5},\,\,\, u_{22}=\frac{B\sin\theta}{r^5}.
\end{equation}
The corresponding matrix potential for the coupled system is
\begin{eqnarray}
\label{5.9}
{\bf V} = \left(\begin{array}{cc}
\frac{(A+B)\sin\theta}{2r^5}-\frac{4}{r^2(r+\cos\theta)^2}
& \frac{(A-B)\sin\theta}{2r^5} \\ \\
-\frac{(A-B)\sin\theta}{2r^5}
&-\frac{(A+B)\sin\theta}{2r^5}+\frac{4}{r^2(r+\cos\theta)^2}
\end{array}
\right).
\end{eqnarray}

\setcounter{equation}{0}
\section{${\bf C}_2$ and ${\bf C}_3$ have Orthogonal Columns}

In this case we assume that ${\bf C}_2$ and ${\bf C}_3$ have the following 
structure
\begin{eqnarray}
\label{3.37}
{\bf C}_2 = \left(\begin{array}{cc}
f_{11}(x,y) &f_{22}(x,y)  \\
-f_{11}(x,y) &f_{22}(x,y) \\
\end{array}
\right),\,\,\,
{\bf C}_3 = \left(\begin{array}{cc}
g_{11}(x,y) &g_{22}(x,y)  \\
-g_{11}(x,y) &g_{22}(x,y) \\
\end{array}
\right).
\end{eqnarray}
where $f_{ii}$ and $g_{ii}$ are complex conjugate. We assume also that 
the desired form of  ${\bf U}$ is given by (\ref{3.2}). Using (\ref{3.3})
to define ${\bf C}_1$, equation (\ref{2.13}) yields after some algebra 
\begin{equation}
\label{3.4}
f_{11}(x, y)\frac{\partial c_{k1}}{\partial y}-
g_{11}(x, y)\frac{\partial c_{k1}}{\partial x}=0,\,\,\, k=1,2,
\end{equation}
\begin{equation}
\label{3.5}
f_{22}(x, y)\frac{\partial c_{k2}}{\partial y}-
g_{22}(x, y)\frac{\partial c_{k2}}{\partial x}=0,\,\,\, k=1,2.
\end{equation}
This leads us to consider the following equations
\begin{equation}
\label{3.6}
f_{11}(x, y)\,dx +g_{11}(x, y)\,dy=0,\,\,\, 
f_{22}(x, y)\,dx +g_{22}(x, y)\,dy=0.
\end{equation}
Although these equations are not exact an integrating factor for these
equations is given by
$$
\frac{1}{f_{11}^2(x,y)+g_{11}^2(x,y)}
$$
and
$$
\frac{1}{f_{22}^2(x,y)+g_{22}^2(x,y)}
$$
respectively. (This fact follows from Cauchy-Riemann equations for
${\bf C}_2$ and ${\bf C}_3$). The general solution of these equations
can be expressed therefore by the standard formulas
\begin{equation}
\label{3.7}
w_i(x,y)=\int_{x_0}^{x}\frac{f_{ii}(x,y)}{f_{ii}^2(x,y)+g_{ii}^2(x,y)}\,dx +
\int_{y_0}^{y}\frac{g_{ii}(x_0,y)}{f_{ii}(x_0,y)^2+g_{ii}(x_0,y)^2}\,dy,
\,\,\, i=1,2.
\end{equation}
It follows then that the general solution for the entries of the matrix
${\bf C}_1$ is
\begin{equation}
\label{3.8}
c_{11}=F_{11}(w_1),\,\,\, c_{12}=F_{12}(w_2),\,\,\, c_{21}=F_{21}(w_1),
\,\,\, c_{22}=F_{22}(w_2).
\end{equation}
However in this case we shall let $F_{12}(w_2)=F_{22}(w_2)$ and 
$F_{21}(w_1)=-F_{11}(w_1)$ i.e the matrix ${\bf C}_1$ is of the form
\begin{eqnarray}
\label{3.38}
{\bf C}_1 = \left(\begin{array}{cc}
F_{11}(w_1) &F_{22}(w_2)  \\
-F_{11}(w_1) &F_{22}(w_2) \\
\end{array}
\right).
\end{eqnarray}
The resulting matrix ${\bf E}$ is the form
\begin{eqnarray}
\label{3.39}
{\bf E} = \left(\begin{array}{cc}
h_{11}(x,y) &h_{22}(x,y)  \\
-h_{11}(x,y) &h_{22}(x,y) \\
\end{array}
\right).
\end{eqnarray}
where $h_{ii}$ are defined by 
\begin{equation}
\label{3.40}
h_{11}=\frac{F_{11}(w_1)^{\prime\prime} -
2F_{11}(w_1)F_{11}(w_1)^{\prime}}{(f_{11}^2+g_{11}^2)},
\end{equation}
\begin{equation}
\label{3.40a}
h_{22}=\frac{F_{22}(w_2)^{\prime\prime} -
2F_{22}(w_2)F_{22}(w_2)^{\prime}}{(f_{22}^2+g_{22}^2)}.
\end{equation}
Substituting these results in (\ref{2.14}) we obtain (only) two independent 
equations for $u_{ii}$.
\begin{equation}
\label{3.41}
g_{ii}\frac{\partial u_{ii}}{\partial y}+
f_{ii}\frac{\partial u_{ii}}{\partial x}+
2\frac{\partial g_{ii}}{\partial y} u_{ii}= h_{ii},\,\,\,i=1,2. 
\end{equation}
Using (\ref{2.12}) we find that the elements of the matrix ${\bf V}$ are
\begin{equation}
\label{3.42}
{\bf V}_{11}={\bf V}_{22}=
\frac{1}{2}(u_{11} + u_{22})-\frac{F_{11}^{\prime}}{(f_{11}^2+g_{11}^2)}
-\frac{F_{22}^{\prime}}{(f_{22}^2+g_{22}^2)},
\end{equation}
\begin{equation}
\label{3.43}
{\bf V}_{12}={\bf V}_{21}=
\frac{1}{2}(u_{22} - u_{11})+\frac{F_{11}^{\prime}}{(f_{11}^2+g_{11}^2)}
-\frac{F_{22}^{\prime}}{(f_{22}^2+g_{22}^2)}.
\end{equation}

We now consider the particular case where
\begin{eqnarray}
\label{5.11}
{\bf C}_2 = \left(\begin{array}{cc}
aRe(z^n) &bRe(z^m) \\
-aRe(z^n) &bRe(z^m) \\
\end{array}
\right),\,\,\,
{\bf C}_3 = \left(\begin{array}{cc}
aIm(z^n) &bIm(z^m)  \\
-aIm(z^n) &bIm(z^m) \\
\end{array}
\right).
\end{eqnarray}
where $a,b$ are real and $n,m > 0$, $m\ne n$. (When $n=m$
this case reduces to the situation discussed in the previous section).  

As in the previous section it is expedient to work this example using 
polar coordinates. Eqs (\ref{3.4}) reduces to (\ref{3.26}) and (\ref{3.5})
reduces to the same equation with $n$ being replaced by $m$. The general
form of the matrix $C_1$ is given by (\ref{3.38}) where
\begin{equation}
\label{5.12}
w_1=\frac{\cos((n-1)\theta)}{r^{n-1}},\,\,\,
w_2=\frac{\cos((m-1)\theta)}{r^{m-1}}.
\end{equation}
Setting $h_{11}=h_{22}=0$ we obtain from (\ref{3.40}),(\ref{3.40a})
that
\begin{equation}
\label{5.13}
F_{11}= \frac{\tan(\frac{w_1+c_2}{c_1})}{c_1},\,\,\, or\,\, F_{11}=constant,
\end{equation}
\begin{equation}
\label{5.13a}
F_{22}= \frac{\tan(\frac{w_2+c_4}{c_3})}{c_3},\,\,\, or\,\, F_{22}=constant,
\end{equation}
where $c_i,\,\,i=1\ldots 4$ are constants. Eqs (\ref{3.41}) for $i=1,2$
then reduce to (\ref{3.31}) (with $n$ being replaced by $m$ for $i=2$)
and therefore (using (\ref{3.32}))
\begin{equation}
\label{5.14}
u_{11}= \frac{1}{r^{2n}}H_{11}\left(\frac{\sin((n-1)\theta)}{r^{n-1}}\right),
\end{equation}
\begin{equation}
\label{5.15}
u_{22}= \frac{1}{r^{2m}}H_{22}\left(\frac{\sin((m-1)\theta)}{r^{m-1}}\right),
\end{equation}
where $H_{11},\,H_{22}$ are smooth functions of the indicated variables.
The explicit general form of the matrix potential $V$ for the coupled system 
in this case is given by (\ref{3.42})-(\ref{3.43}).

If we let $n=2$, $m=1$, $F_{ii}=constant,\,i=1,2$ and choose
\begin{equation}
\label{5.16}
H_{11}=\frac{A\sin\theta}{r},\,\,\, H_{22}=B
\end{equation}
where $A,B$ are constants then the matrix potential $V$ is given by 
\begin{eqnarray}
\label{5.17}
{\bf V} = \frac{1}{2}\left(\begin{array}{cc}
\frac{A\sin\theta}{r^5}+\frac{B}{r^2}
&\frac{B}{r^2}-\frac{A\sin\theta}{r^5} \\ \\
\frac{B}{r^2}-\frac{A\sin\theta}{r^5}
&\frac{A\sin\theta}{r^5}+\frac{B}{r^2}
\end{array}
\right).
\end{eqnarray}

\section{Summary and Conclusions.}

In this paper we showed that Darboux transformations in two dimensions
can be applied to decouple systems of PDES. We demonstrated also
the close affinity of these transformations to complex analytic functions.
Although we were unable to solve (\ref{2.13})-(\ref{2.14}) in general
we were able to provide explicit solutions for three classes of 
equations. Even if the decoupled system is not integrable analytically
its numerical solution is less demanding computationally. Furthermore
the decoupled system might provide insights about the solutions 
of the original system which are obvious directly.

\newpage
\section*{References}

\begin{itemize}

\item[1] C. Gu, H. Chaohao and Z. Zhou - Darboux Transformations in 
Integrable Systems, Springer, New-York (2005)

\item[2] C. Rogers W. K. Schief - Backlund and Darboux Transformations:
Geometry and Modern Applications in Soliton Theory,
Cambridge Univ. Press, Cambridge(UK) (2005)

\item[3] M. Humi - Separation of coupled systems of differential equations by 
Darboux transformation.  J. of Physics A  {\bf 18}, p. 1085 (1985).

\item[4]  M. Humi - Darboux transformations for Schroedinger equations 
in two variables, J. Math. Phys. {\bf 46}, 083515 (2005) (8 pages).

\item[5] Boyan Sirakov, Sérgio H. M. Soares-
Soliton solutions to systems of coupled Schrödinger equations of
Hamiltonian type, Trans. Amer. Math. Soc. 362,  5729-5744 (2010).

\item[6] Samsonov, Boris F., Pecheritsin, A. A - Chains of Darboux 
transformations for the matrix Schrödinger equation, Journal of Physics 
A 37, pp. 239-250 (2004)

\item[7] F. Magri and J. P. Zubelli - Differential equations in the 
spectral parameter, Darboux transformations and a hierarchy of
master symmetries for KdV, Comm. Math. Phys.  {\bf 141}, p.329-335 (1991).

\item[8] A. Gonzalez-Lopez and N. Kamran - The Multidimensional Darboux 
Transformation, J. Geom. Phys. {\bf 26}, p. 202-226 (1998).

\item[9] A. Adiloglu Nabiev- On a fundamental system of solutions of 
the matrix Schrödinger equation with a polynomial energy-dependent potential
Mathematical Methods in the Applied Sciences 33, pp. 1372-1383 (2010).

\item[10]  D. Adamova , J. Horejsi and I. Ulehla - The Atkinson-Prufer 
transformation and the eigenvalue problem for coupled systems of 
Schrodinger equations, J. Phys. A: Math. Gen. 17 p. 2621 (1984)

\item[11] H. Amann and P. Quittner - A nodal theorem for coupled systems 
of Schrodinger equations and the number of bound states, J. Math. Phys.
36 p.4553 (1995).

\item[12] A.A. Suzko- Intertwining technique for the matrix Schrodinger 
equation, Physics Letters A, Volume 335, pp. 88-102 (2005)

\item[13] M. Humi - Factorization of systems of differential equations
     J. Math. Phys. {\bf 27}, p.76 (1986).

\item[14] M. Humi - Factorization of separable partial differential 
equations, J. Phys. A {\bf 20}, p.4577 (1987).

\item[15] A.A. Andrianov, M.V. Ioffe and D.N. Nishnianidze - Polynomial 
SUSY in Quantum Mechanics and Second Derivative Darboux Transformation
Phys.Lett. A {\bf 201}, p. 103 (1995).

\item[16] F. Finkel, A. Gonzalez-Lopez and N. Kamran - On form-preserving
transformations for the time dependent Schroedinger equation, J. Math. Phys. 
{\bf 40}, p.3268-3274 (1999).

\item[17] A.B. Balantekin - Algebraic approach to shape invariance,
Physical Review A {\bf 57}, pp.4188-4191 (1998).

\end{itemize}

\end{document}